# Kaon structure in the nuclear medium within the light front approach

G. H. S. O. Yabusaki,[1,2,3] J. P. B. C. de Melo,[2] K. Tsushima,[2] T. Frederico,[3] and W. de Paula[3]

[1]*Universidade São Judas Tadeu, 03166-000, São Paulo, Brazil*
[2]*Laboratório de Física Teórica e Computacional-LFTC,
Universidade Cruzeiro do Sul and Universidade Cidade de São Paulo, 01506-000, São Paulo, Brazil*
[3]*Instituto Tecnológico de Aeronáutica, DCTA, 12228-900, São José dos Campos, Brazil*

We study the properties of the charged kaon in symmetric nuclear matter using a Bethe-Salpeter amplitude to model the quark-antiquark bound state, which is well constrained by previous studies of its vacuum properties. The electromagnetic form factor, charge radius, decay constant, and the light-front valence component probability are investigated in symmetric nuclear matter. In order to describe the constituent up and antistrange quarks in nuclear matter, we adopt the "quark-meson coupling (QMC) model," which has been widely applied to various hadronic and nuclear phenomena in the nuclear medium.

## I. INTRODUCTION

The main purpose of this work is to investigate the in-medium modification of the $K^+$-meson properties in symmetric nuclear matter in a light-front constituent quark model with the in-medium input of the quark-meson coupling (QMC) model [1–3], where the $K^+$-meson model [4] is adjusted to provide the best description of the $K^+$-meson data in vacuum. Similar approaches were also used for the pion [5–8]. Furthermore, we mention that their static and dynamical properties have also been investigated theoretically and experimentally [5,9–28].

The kaon has not received as much attention compared to the pion, however they prove to be important as they have an abundant source of information about the nature of fundamental interactions essential in establishing the standard model of particle physics, such as the quark model of hadrons [14,29–34]. The kaon structure and their dynamics are an important tool to understand the electroweak interactions and $CP$ violation [30].

The theoretical framework we adopt is the light-front approach [35,36], more specifically, we use a symmetric vertex model for $K^+$-meson bound state $|u\bar{s}\rangle$ in order to model the Bethe-Salpeter amplitude [4,5,37].

An important additional information about the meson's internal structure can be inferred from their valence-quark distribution functions. With respect to the description of bound states on the light-cone approach, a detailed review of hadronic wave functions in QCD based models can be found in Ref. [35]. The light-front plus component of the electromagnetic current $J^+$, has been successfully used to calculate elastic electromagnetic form factors, charge radii, and also the weak decay constants for pseudoscalar particles with the light-front approach [4,5,9–11,30,33]. Using the symmetric $K - q\bar{q}$ vertex model [5], the components of the current are conveniently obtained in the Drell-Yan frame, where the light-front bound state wave functions are defined on the hypersurface $x^0 + x^3 = 0$ and are covariant under kinematical boosts due to the stability of Fock-state decomposition [35,38], and it was possible to analyse the partial contributions of each particle to the composition of the complete form factor. In the present work, we consider the symmetric vertex function to optimize and unify the parameter set to simultaneously calculate the electromagnetic form factor, root-mean-square charge radius, decay constant, and probability of the kaon valence component $\eta$ [4,5]. Our numerical results are compared with experimental data in vacuum up to $Q^2 \approx 0.10 \text{ GeV}^2$ to explore the validity of the model, where $Q^2 = -q^2 > 0$, with $q$ being the four-momentum transfer.

This work is organized as follows. In Sec. II we briefly review the QMC model focusing on the properties of constituent up and strange quarks and the kaon vertex in symmetric nuclear matter. The expressions for the in-medium electromagnetic form factor of the kaon are discussed in Sec. III, while the results for the in-medium kaon properties, electromagnetic form factor and partial contribution of charge, decay constant and probability of the kaon valence component $\eta$ are presented in Sec. IV. Finally, Sec. V is devoted to a summary and discussions with the perspective of applying the model to other mesons.

## II. QUARKS IN NUCLEAR MATTER

The QMC model was developed in 1988 by Guichon [1] with the MIT bag model, and a similar model by Frederico *et al.* in 1989 [2] with a confining harmonic potential. In both approaches it is possible to describe nuclear matter properties based on the quark degrees of freedom. The model has been successfully applied for various studies of finite (hyper)nuclei [39–43] as well as the hadron properties in a nuclear medium (see Ref. [3] for a review).

The medium effects arise through the self-consistent coupling of phenomenological isoscalar-Lorentz-scalar ($\sigma$), isoscalar-Lorentz-vector ($\omega$), and isovector-Lorentz-vector ($\rho$) meson fields directly to the confined light flavor $u$ and $d$ valence quarks, rather than to the nucleons and the heavier flavor quarks. As a result, the internal structure of the bound nucleon is modified by the surrounding nuclear medium with respect to the free nucleon. The effective Lagrangian density for a uniform, spin saturated and isospin-symmetric nuclear system (symmetric nuclear matter) at the hadronic level is given by [1–3,39–43]

$$\mathcal{L} = \bar{\psi}[i\gamma \cdot \partial - m_N^*(\hat{\sigma}) - g_\omega \hat{\omega}^\mu \gamma_\mu]\psi + \mathcal{L}_{\text{meson}}, \quad (1)$$

where $\psi$, $\hat{\sigma}$ and $\hat{\omega}$ are respectively the nucleon, Lorentz-scalar isoscalar $\sigma$, and Lorentz-vector isoscalar $\omega$ field operators, with $g_\omega$ being the nucleon-$\omega$ coupling constant, while the nucleon-$\sigma$ effective coupling which depends on the $\hat{\sigma}$ (or nuclear density) is defined by,

$$m_N^*(\hat{\sigma}) = m_N - g_\sigma(\hat{\sigma})\hat{\sigma}, \quad (2)$$

which defines the $\sigma$-field dependent coupling constant, $g_\sigma(\hat{\sigma})$. All the important nuclear many-body dynamics, including three-body nucleon force modeled at the quark level, will effectively be condensed in $g_\sigma(\hat{\sigma})$. Solving the Dirac equations for the up and down quarks in the nuclear medium with the same mean fields (mean values) $\sigma$ and $\omega$, which act on the bound nucleon self-consistently based on the Lagrangian, we obtain the effective $\sigma$-dependent coupling $g_\sigma(\sigma)$ at the hadronic level [1,2,39–43]. The free meson Lagrangian density is given by

$$\mathcal{L}_{\text{meson}} = \frac{1}{2}\left(\partial_\mu \hat{\sigma} \partial^\mu \hat{\sigma} - m_\sigma^2 \hat{\sigma}^2\right) - \frac{1}{2}\partial_\mu \hat{\omega}_\nu\left(\partial^\mu \hat{\omega}^\nu - \partial^\nu \hat{\omega}^\mu\right) + \frac{1}{2}m_\omega^2 \hat{\omega}^\mu \hat{\omega}_\mu, \quad (3)$$

where we neglected the isospin-dependent Lorentz-vector isovector $\rho$-meson field, since we consider isospin-symmetric nuclear matter within the Hartree mean-field approximation. In this case the mean value of the $\rho$-mean field becomes zero and there is no need to consider its possible contributions due to the $\rho$-Fock (exchange) terms.

In the sequence we adopt the nuclear matter rest frame. Then, the nucleon density $\rho$, the nucleon Fermi momentum $k_F$, the nucleon scalar density $\rho_s$, and the effective nucleon mass $m_N^*$ are related by,

$$\rho = \frac{4}{(2\pi)^3}\int d^3k \theta(k_f - |\mathbf{k}|) = \frac{2k_f^3}{3\pi^2},$$
$$\rho_s = \frac{4}{(2\pi)^3}\int d^3k \theta(k_f - |\mathbf{k}|)\frac{m_N^*(\sigma)}{\sqrt{m_N^{*2} + \mathbf{k}^2}}, \quad (4)$$

where $m_N^*(\hat{\sigma})$ [see Eq. (2)], is the value of the effective nucleon mass at a given density, calculated by the QMC model [1,2,39–44] [See also Eq. (9)]. The Dirac equations for the light quarks, light antiquarks, strange and heavy quarks, and strange and heavy antiquarks ($q = u$ or $d$ and $Q = s$, $c$ or $b$) in the bag of hadron $h$ in nuclear matter at the position $x = (t, \mathbf{r})$ with $|\mathbf{r}| \leq$ bag radius, are given by [3],

$$\left[i\gamma \cdot \partial_x - (m_q - V_\sigma^q) \mp \gamma_0\left(V_\omega^q + \frac{1}{2}V_\rho^q\right)\right]\begin{pmatrix}\psi_u(x)\\\psi_{\bar{u}}(x)\end{pmatrix} = 0, \quad (5)$$

$$\left[i\gamma \cdot \partial_x - (m_q - V_\sigma^q) \mp \gamma_0\left(V_\omega^q + \frac{1}{2}V_\rho^q\right)\right]\begin{pmatrix}\psi_d(x)\\\psi_{\bar{d}}(x)\end{pmatrix} = 0, \quad (6)$$

$$[i\gamma \cdot \partial_x - m_Q]\begin{pmatrix}\psi_Q(x)\\\psi_{\bar{Q}}(x)\end{pmatrix} = 0, \quad (7)$$

where we have neglected the Coulomb force as usual, since the nuclear matter properties are due to the strong interaction, and we assume SU(2) symmetry for the light quarks, $m_q = m_u = m_d$, and define $m_q^* = m_q - V_\sigma^q = m_u^* = m_{\bar{u}}^*$, but $m_{\bar{q}} = m_s^* = m_{\bar{s}}$. In symmetric nuclear matter, the isospin dependent $\rho$-meson mean field in Hartree approximation yields $V_\rho^q = 0$ in Eqs. (5) and (6), so we ignore it hereafter.

The constant mean-field potentials in nuclear matter are defined by $V_\sigma^q \equiv g_\sigma^q \sigma = g_\sigma^q \langle\sigma\rangle$ and $V_\omega^q \equiv g_\omega^q \omega = g_\omega^q \delta^{\mu,0}\langle\omega^\mu\rangle$, with $g_\sigma^q$ and $g_\omega^q$ being the corresponding quark-meson coupling constants, and the quantities inside the brackets stand for taking expectation values by the nuclear matter ground state [3]. Note that, since the velocity averages to be zero in the rest frame of nuclear matter, the mean vector source due to the quark fields as well, $\langle\bar{\psi_q}\vec{\gamma}\psi_q\rangle = 0$. Thus we may just keep the term proportional to $\gamma^0$ in Eqs. (5) and (6). The normalized, static solution for the ground state quarks or antiquarks with flavor $f$ in the hadron $h$, may be written as $\psi_f(x) = N_f e^{-i\epsilon_f t/R_h^*}\psi_f(\vec{r})$, where $N_f$ and $\psi_f(\vec{r})$ are the normalization factor and corresponding spin and spatial part of the wave function. The bag radius in medium for a hadron $h$, $R_h^*$, is determined through the stability condition for the mass of the hadron against the variation of the bag radius [3] [See also Eq. (9)]. The eigenenergies in units of $1/R_h^*$ are given by

$$\begin{pmatrix} \epsilon_u \\ \epsilon_{\bar{u}} \end{pmatrix} = \Omega^* \pm R_h^* \left( V_\omega^q + \frac{1}{2} V_\rho^q \right),$$
$$\begin{pmatrix} \epsilon_d \\ \epsilon_{\bar{d}} \end{pmatrix} = \Omega^* \pm R_h^* \left( V_\omega^q - \frac{1}{2} V_\rho^q \right),$$
$$\epsilon_Q = \epsilon_{\bar{Q}} = \Omega_Q^*. \quad (8)$$

The hadron masses in a nuclear medium $m_h^*$ (free mass $m_h$), are calculated by

$$m_h^* = \sum_{j=q,\bar{q},Q,\bar{Q}} \frac{n_j \Omega_j^* - Z_h}{R_h^*} + \frac{4}{3}\pi R_h^{*3} B,$$
$$\left. \frac{dm_h^*}{dR_h^*} \right|_{R_h = R_h^*} = 0, \quad (9)$$

where $\Omega_q^* = \Omega_{\bar{q}}^* = [x_q^2 + (R_h^* m_q^*)^2]^{1/2}$, with $m_q^* = m_q - g_\sigma^q \sigma$, $\Omega_Q^* = \Omega_{\bar{Q}}^* = [x_Q^2 + (R_h^* m_Q)^2]^{1/2}$, and $x_{q,Q}$ being the lowest bag eigenfrequencies. $n_q(n_{\bar{q}})$ and $n_Q(n_{\bar{Q}})$ are the quark (antiquark) numbers for the quark flavors $q$ and $Q$, respectively. Note that, when the hadron $h$ contains at least one light quark (antiquark) $R_h^* \neq R_h$ and thus $\Omega_Q^* \neq \Omega_Q$. The MIT bag quantities, $Z_h$, $B$, $x_{q,Q}$ and $m_{q,Q}$ are the parameters for the sum of the center of mass and gluon fluctuation effects, bag constant, lowest eigenvalues for the quarks $q$ or $Q$, respectively, and the corresponding current quark masses. We will also give the definition of incompressibility (K), which is a measure of the resistance of atomic nuclei to compression, describing the rigidity of the nuclear matter (see Eq. (14) [1,45]). $Z_N$ and $B$ are fixed by fitting the nucleon (the hadron) mass in free space. (See Table I the nucleon case.) For the nucleon $h = N$ case in the above, the lowest, positive bag eigenfunction is given by

$$q(t,\vec{r}) = \frac{\mathcal{N}}{\sqrt{4\pi}} e^{-i\epsilon_q t/R_N^*} \begin{pmatrix} j_0(xr/R_N^*) \\ i\beta_q^* \vec{\sigma} \cdot \hat{v} j_1(xr/R_N^*) \end{pmatrix} \theta(R_N^* - r) \chi_m, \quad (10)$$

TABLE I. Coupling constants, the parameter $Z_N$, bag constant $B$ (in $B^{1/4}$), and calculated properties for symmetric nuclear matter at normal nuclear matter density $\rho_0 = 0.15$ fm$^{-3}$, for $m_q = 5$ and 220 MeV. The effective nucleon mass, $m_N^*$, and the nuclear incompressibility, $K$, are quoted in MeV (the free nucleon bag radius used is $R_N = 0.8$ fm, the standard input value in the QMC model [3]).

| $m_q$ (MeV) | $g_\sigma^2/4\pi$ | $g_\omega^2/4\pi$ | $m_N^*$ | $K$ | $Z_N$ | $B^{1/4}$ (MeV) |
|---|---|---|---|---|---|---|
| 5 | 5.39 | 5.30 | 754.6 | 279.3 | 3.295 | 170 |
| 220 | 6.40 | 7.57 | 698.6 | 320.9 | 4.327 | 148 |

with $r = |\vec{r}|$ and $\chi_m$ the spin function and

$$\Omega_q^* = \sqrt{x^2 + (m_q^* R_N^*)^2},$$
$$\beta_q^* = \sqrt{\frac{\Omega_q^* - m_q^* R_N^*}{\Omega_q^* + m_q^* R_N^*}},$$
$$\mathcal{N}^{-2} = \frac{2R_N^{*3} j_0^2(x)\left[\Omega_q^*(\Omega_q^* - 1) + \frac{m_q^* R_N^*}{2}\right]}{x^2}, \quad (11)$$

where $x$ is the eigenvalue for the lowest mode, which satisfies the boundary condition at the bag surface, $j_0(x) = \beta_q^* j_1(x)$. The same meson mean fields $\sigma$ and $\omega$ for the quarks satisfy the following equations at the nucleon level self-consistently:

$$\sigma = \frac{g_\sigma}{m_\sigma^2} C_N(\sigma) \frac{4}{(2\pi)^3} \int d\vec{k}\, \theta(k_F - |\mathbf{k}|) \frac{m_N^*}{\sqrt{m_N^{*2} + \mathbf{k}^2}},$$
$$\omega = \frac{g_\omega \rho}{m_\omega^2},$$
$$C_N(\sigma) = \frac{-1}{g_\sigma(\sigma = 0)} \left[ \frac{\partial m_N^*(\sigma)}{\partial \sigma} \right], \quad (12)$$

where $C_N(\sigma)$ is the constant value of the scalar density ratios [1,2,39–43].

Because of the underlying quark structure of the nucleon used to calculate $m_N^*(\sigma)$ in the nuclear medium [see Eq. (9) with $h = N$], $C_N(\sigma)$ gets $\sigma$ dependence, whereas the usual pointlike nucleon based model yields unity, $C_N(\sigma) = 1$. It is this $C_N(\sigma)$ or $g_\sigma(\sigma)$ that gives a new saturation mechanism in the QMC model. Without an explicit introduction of nonlinear couplings of the meson fields in the Lagrangian density at the nucleon and meson level, the standard QMC model yields the nuclear incompressibility of $K \simeq 280$ MeV, which is in contrast to a naive version of quantum hydrodynamics (QHD) [46], also called the pointlike nucleon model of nuclear matter, which results in a much larger value, $K \simeq 500$ MeV; the empirically extracted value falls in the range $K = 200$–$300$ MeV [45]. Once the self-consistency equation for the $\sigma$, Eq. (12), has been solved, one can evaluate the total energy per nucleon, which is shown in Eq. (13) and Fig. 1,

$$W = E^{\text{tot}}/A = \frac{4}{(2\pi)^3 \rho} \int d^3k\, \theta(k_F - |\mathbf{k}|)$$
$$\times \sqrt{m_N^{*2} + \vec{k}^2} + \frac{m_\sigma^2 \sigma^2}{2\rho} + \frac{g_\omega^2 \rho}{2m_\omega^2}. \quad (13)$$

The nuclear incompressibility K is evaluated according to the equation below [1,45],

$$K = 9\rho_0 \left. \frac{\partial^2 W}{\partial \rho^2} \right|_{(\rho=\rho_0)}, \quad (14)$$

where $W = E/A$ is the energy per particle in Eq. (13).

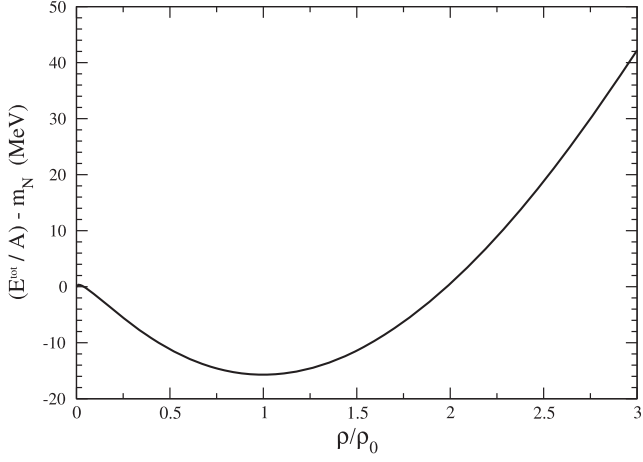

FIG. 1. Negative of the binding energy per nucleon $(E^{\text{tot}}/A) - m_N$ for symmetric nuclear matter obtained via the vacuum up and down quark mass, $m_q = 220$ MeV. At the saturation point $\rho_0 = 0.15$ fm$^{-3}$, the value is fitted to $-15.7$ MeV.

We then determine the coupling constants, $g_\sigma$ and $g_\omega$, so as to fit the binding energy of 15.7 MeV at the saturation density $\rho_0 = 0.15$ fm$^{-3}$ ($k_F^0 = 1.305$ fm$^{-1}$) for symmetric nuclear matter.

The kaon model we adopt here (see the Refs. [4,5]), uses a vacuum constituent quark mass, $m_u = 220$ MeV and $m_{\bar{s}} = 508$ MeV [47], in order to well reproduce the electromagnetic form factor and decay constant data. Therefore, to be consistent with this kaon model, our nuclear matter is built with the same vacuum mass. The corresponding coupling constants and some calculated properties for symmetric nuclear matter at the saturation density, with the standard values of $m_\sigma = 550$ MeV and $m_\omega = 783$ MeV, are listed in Table I.

For comparison, we also give the corresponding quantities calculated in the standard QMC model with a vacuum quark mass of $m_q = 5$ MeV as shown in Ref. [3]. Thus we have obtained the necessary properties of the light-flavor constituent quarks in symmetric nuclear matter with the empirically accepted data for a vacuum mass of $m_u = 220$ MeV and strange quark mass $m_{\bar{s}} = 508$ MeV [47]; namely, the density dependence of the effective mass (scalar potential) and vector potential. The same in-medium constituent light quark properties will be used as input as already used to describe the pion immersed in symmetric nuclear matter [6].

We will use the $|V_\omega^K|$ in Fig. 3 correspondingly as $|V_\omega^q|$ in the calculation and the kaon mass, $m_K^* = m_h^*$, being calculated by Eq. (9) with $R_h = R_K$ and $Z_h = Z_K$.

In Fig. 2 we show the results for the effective mass of the constituent light quark and potentials versus nuclear density. In Fig. 3 shown is the kaon bound state effective mass and potentials versus nuclear density, and in Fig. 4 we show the effective mass of the nucleon, $m_N^*$ in symmetric nuclear matter. Note that the total potential of $K^+$ at $\rho_0$ is slight repulsion of about 40 MeV as empirically extracted and practiced [44,48,49]. However, according to Fig. 2, the effective "constituent quark mass" becomes negative at $\rho \simeq 1.8\rho_0$. Thus, the present constituent quark kaon model in symmetric nuclear matter may be valid in the range $\rho = [0, 1.8\rho_0]$. The consequence of this model validity will be reflected on some results later, and one must take the results with caution for $\rho > 1.8\rho_0$. (Explicitly some results will be shown in Figs. 9 and 10). In the sequence, the symmetric vertex model for the kaon with the potentials of the nuclear medium will be shown.

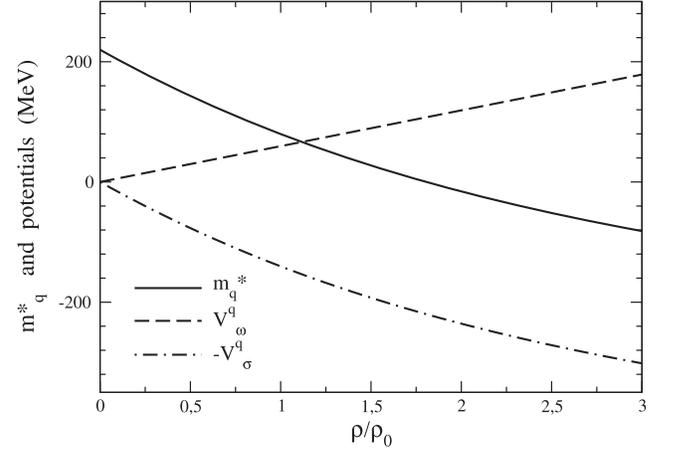

FIG. 2. Effective light quark masses and the potentials in symmetric nuclear matter.

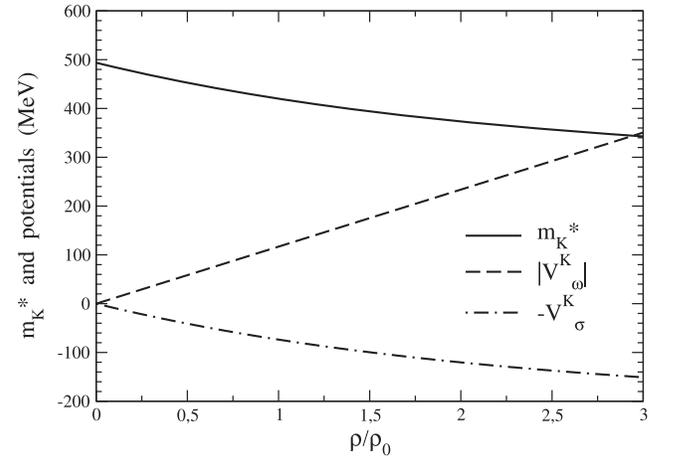

FIG. 3. Kaon effective mass and potentials versus the nuclear density, where total potential at $\rho_0$ gives a slight repulsion of about 40 MeV.

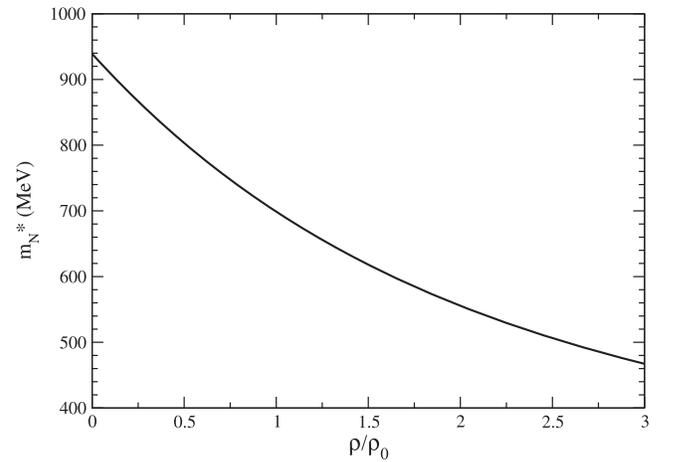

FIG. 4. Nucleon effective mass, $m_N^*$ versus nuclear density.

## III. THE MODEL

The electromagnetic current for a $K^+$-meson ($|u\bar{s}\rangle$ bound state), is calculated in one-loop approximation (triangle diagram shown in Fig. 5), modeling the Bethe-Salpeter amplitude through a symmetric vertex function in momentum space with a pseudoscalar coupling between $K^+$ and quarks. This coupling is given by the effective Lagrangian in vacuum [5,33,37],

$$\mathcal{L}_{\text{eff}} = -i\frac{\hat{m}}{f_{K^+}}\bar{q}\frac{1}{\sqrt{2}}(\lambda_4 + i\lambda_5)\gamma^5 q\frac{1}{\sqrt{2}}(\phi_4 - i\phi_5)\Lambda, \quad (15)$$

here, $q = (u, d, s)^T$ and $K^+ = \frac{1}{\sqrt{2}}(\phi_4 - i\phi_5)$, $\hat{m} = \frac{m_u + m_{\bar{s}}}{2}$, $\Lambda$ the symmetric vertex function, and $f_{K^+}$ the $K^+$ decay constant. With the Lagrangian (15) we will work in Hartree mean field approximation and the modifications will enter as the shift of the light-quark momentum in light front approach via $m_K \to m_K^*$ and $P^+ = P^0 + P^3 \to P^{*+} = P^{*0} + P^{*3} + V = \sqrt{m_K^2 + \vec{q}^2/4} + P^{*3} + V$ due to the vector potential $V$, and in the Lorentz-scalar part through the Lorentz-scalar potential $V_\sigma^q$ as $m_u \to m_u^* = m_u - V_\sigma^q$ [3,6] and $m_{\bar{s}} \to m_{\bar{s}}^* = m_{\bar{s}}$ based on the QMC model [see Eq. (6)] [3]. The QMC model has been applied to many nuclear and hadronic phenomena in a nuclear medium with success, and the inputs in vacuum as well as quantities calculated in-medium shown in Table I are adopted in the present model to describe the effects of nuclear medium as seen in the previous section. The electromagnetic current for the $K^+$ with the plus-component, is obtained from the covariant expression Eq. (16) corresponding to the vacuum triangle diagram in Fig. 5.

In the vacuum the electromagnetic current is given by,

$$J^\mu(q^2) = -ie\frac{2\hat{m}^2}{f_{K^+}^2}N_C \int \frac{d^4k}{(2\pi)^4}$$
$$\times \left\{\frac{1}{3}\text{Tr}[S(k, m_u)\gamma^5 S(k - P', m_{\bar{s}})\gamma^\mu S(k - P, m_{\bar{s}})\gamma^5]\right.$$
$$\left.+\frac{2}{3}\text{Tr}[S(k, m_{\bar{s}})\gamma^5 S(k - P', m_u)\gamma^\mu S(k - P, m_u)\gamma^5]\right\}$$
$$\times \Lambda(k, P)\Lambda(k, P'), \quad (16)$$

where the symmetric vertex function is given by [4–6]

$$\Lambda(k, P) = \frac{C}{k^2 - m_R^2 + i\epsilon} + \frac{C}{(P-k)^2 - m_R^2 + i\epsilon}, \quad (17)$$

with $N_C = 3$ being the number of colors in QCD, and

$$S(k - P, m_u) = \frac{1}{(\slashed{k} - \slashed{P}) - m_u + i\epsilon},$$
$$S(k - P, m_{\bar{s}}) = \frac{1}{(\slashed{k} - \slashed{P}) - m_{\bar{s}} + i\epsilon}, \quad (18)$$

are corresponding to the up and antistrange quark propagators in the photon interaction, respectively.

Also, we work in the Breit frame and using light-front coordinates, $k^+ = k^0 + k^3$, $k^- = k^0 - k^3$, and $k^\perp = (k^1, k^2)$, and one has

$$q^+ = -q^- = \sqrt{-q^2}\sin\alpha, \quad q_x = \sqrt{-q^2}\cos\alpha, \quad q_y = 0$$
$$\text{and} \quad q^2 = q^+ q^- - (\vec{q}_\perp)^2, \quad (19)$$

where the Drell-Yan condition $q^+ = 0$ is recovered with $\alpha$ equal zero [5,11,50]. As is well known, the $K^+$-meson form factor can be extracted from the covariant expression below, in the elastic case,

$$F_{K^+}(q^2) = \frac{1}{e(P + P')^\mu}\langle P'|J^\mu|P\rangle. \quad (20)$$

In the light-front approach, besides the valence component of the electromagnetic current, we can have the nonvalence contribution or zero modes; thus, the full electromagnetic form factor is given by [5,11,30,31],

$$F_{K^+}(q^2) = F_{K^+}^{(I)}(q^2, \alpha) + F_{K^+}^{(II)}(q^2, \alpha), \quad (21)$$

where the valence component, $F_{K^+}^{(I)}(q^2, \alpha)$, has the loop integration on $k^-$ light-front energy, constrained by $0 \leq k^+ < P^+$ (see the light-front time-ordered diagram in the left panel of the Fig. 6), and $F_{K^+}^{(II)}(q^2, \alpha)$ has the loop integration on $k^-$ in the interval $P^+ \leq k^+ \leq P'^+$ (see the right panel of the Fig. 6), corresponding to pair production contributions with $q^+ > 0$. We use only the valence component, $F_{K^+}^{(I)}(q^2, \alpha)$, since the nonvalence component goes to zero in the adopted framework (further see Refs. [4,5,11,33] for more details).

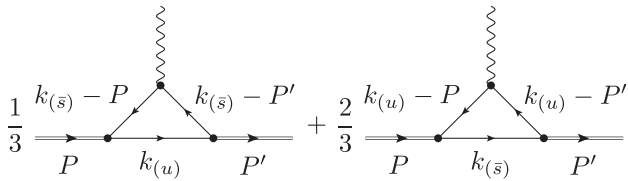

FIG. 5. Feynman diagrams for $K^+$-photon interaction in the vacuum.

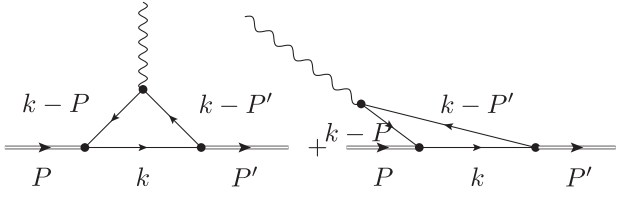

FIG. 6. Feynman diagrams contributing to the elastic electromagnetic form factor for pseudoscalar mesons. Left: the valence component contributions from the electromagnetic current to the electromagnetic form factor. Right: nonvalence contribution for the electromagnetic current with the frame different from the Breit frame with the Drell-Yan condition.

Now we consider the symmetric nuclear matter adapting the initial and final momenta of the composite spin zero bound state defined by energy $P^{*0} = E_K^* = E_K^{\prime *} = \sqrt{m_K^{*2} + \vec{q}^2/4}$ and $\vec{P}_\perp^{\prime} = -\vec{P}_\perp = \frac{\vec{q}}{2}$ where the $q$ conditions using light-front approach are shown by Eq. (19), $\Lambda^*(k, P^*)$ and $\Lambda^*(k, P^{*'})$ using $m_u \to m_u^*$ and $m_{\bar{s}} \to m_{\bar{s}}^*$ as discussed in Sec. II. After the $k^{*-}$ integration for $J^+$ current [see Eq. (16)], a light-front wave function emerges from the symmetric vertex function with the change of variable $x = \frac{k^+}{P^+}$ in medium to $x^* = \frac{k^{*+}}{P^{*+}}$, where $P^{*+} = E_K^* + P^{*3} + V$, $k_u^{*+} = E_u^* + k_u^{*3} + |V_\omega^u|$ for the $u$ quark momenta and $k_{\bar{s}}^{*+} = E_{\bar{s}}^* + k_{\bar{s}}^{*3}$ for the $\bar{s}$ antiquark momenta [5,6,37,44]. We summarize here the light-front model in vacuum for the symmetric vertex function $\Lambda$ ($V = 0$ and $m_R^* = m_R$) for the pseudoscalar bound states. In addition, the $m_R^* = m_R = 600$ MeV is used, the same value as in the vacuum [4,5].

The $K^+$-meson light-front wave function in the symmetric nuclear matter is defined as:

$$\Phi^*(x^*, \vec{k}_\perp) = \frac{P^{*+}}{m_{K^+}^{*2} - \mathcal{M}_0^2} \left[ \frac{\mathcal{N}}{(1-x^*)(m_{K^+}^{*2} - \mathcal{M}_0^2)} + \frac{\mathcal{N}}{x^*(m_{K^+}^{*2} - \mathcal{M}_R^2)} \right] + [u \leftrightarrow \bar{s}], \quad (22)$$

where $\mathcal{N}$ is the normalization factor, $\mathcal{M}_0^2$ is the free mass operator and $\mathcal{M}_R^2$ is a regulator mass function, given below,

$$\mathcal{M}_0^2 = \frac{k_\perp^2 + m_u^{*2}}{x^*} + \frac{(P^* - k)_\perp^2 + m_{\bar{s}}^2}{(1-x^*)} - P_\perp^2, \quad \text{and}$$

$$\mathcal{M}_R^2 = \frac{k_\perp^2 + m_u^{*2}}{x^*} + \frac{(P^* - k)_\perp^2 + m_R^2}{(1-x^*)} - P_\perp^2, \quad (23)$$

with, $[u \leftrightarrow \bar{s}]$ (recall that $m_{\bar{s}}^* = m_{\bar{s}}$). Using only the valence component of the electromagnetic current coming from the Feynman diagram seen in Fig. 5, $F_{K^+}^{(I)}(q^2, \alpha)$, (because the entire production of pairs, $F_{K^+}^{(II)}(q^2, \alpha)$, is null for the adopted reference as previously stated [5,11,30,31]) the elastic electromagnetic form factor for the kaon in nuclear medium, evaluated in the Breit-frame [4–6,11,33] is given by,

$$F_{K^+}^{*(WF)}(q^2) = \frac{1}{2\pi^3(P^{*/+} + P^{*+})}$$
$$\times \int \frac{d^2k_\perp dk^{*+} \theta(k^{*+}) \theta(P^{*+} - k^{*+})}{k^{*+}(P^{*+} - k^{*+})(P^{*/+} - k^{*+})}$$
$$\times \Phi^{*\dagger}(x^*, \vec{k}_\perp) \text{Tr}[\mathcal{O}^+] \Phi^*(x^*, \vec{k}_\perp)$$
$$+ [u \leftrightarrow \bar{s}]. \quad (24)$$

The Eq. (25) represents the Dirac trace $\text{Tr}[\mathcal{O}^+]$, which is calculated with the light-front coordinates in symmetric nuclear matter, and the result is,

$$\text{Tr}[\mathcal{O}^+] = \frac{1}{4} k^{*+} q_\perp^2 - \left( \frac{k_\perp^2 + m_u^{*2}}{k^{*+}} \right) P^{*+} P^{*/+}$$
$$- \left( P^{*/+} k_\perp \cdot P_\perp + P^{*+} k_\perp \cdot P_\perp' \right)$$
$$+ [u \leftrightarrow \bar{s}]. \quad (25)$$

It is also interesting to describe the partial contributions of quarks in medium, for the quark u, $F_{K^+(\bar{s}u\bar{s})}^*$, and the strange antiquark, $F_{K^+(u\bar{s}u)}^*$, to the formation of the full form factor as shown below,

$$F_{K^+(\bar{s}u\bar{s})}^*(q^2) = e_{\bar{s}} \frac{1}{2\pi^3(P^{*/+} + P^{*+})} \int \frac{d^2k_\perp dk^{*+} \theta(k^{*+}) \theta(P^{*+} - k^{*+})}{k^{*+}(P^{*+} - k^{*+})(P^{*/+} - k^{*+})} \Psi^{*\dagger}(x^*, \vec{k}_\perp) \text{Tr}[\mathcal{O}^+]_{\bar{s}u\bar{s}} \Psi^*(x^*, \vec{k}_\perp), \quad (26)$$

$$F_{K^+(u\bar{s}u)}^*(q^2) = e_u \frac{1}{2\pi^3(P^{*/+} + P^{*+})} \int \frac{d^2k_\perp dk^{*+} \theta(k^{*+}) \theta(P^{*+} - k^{*+})}{k^{*+}(P^{*+} - k^{*+})(P^{*/+} - k^{*+})} \Psi^{*\dagger}(x^*, \vec{k}_\perp) \text{Tr}[\mathcal{O}^+]_{u\bar{s}u} \Psi^*(x^*, \vec{k}_\perp), \quad (27)$$

with the values of $e_u = 2/3$ and $e_{\bar{s}} = 1/3$, referring to the charges of the quarks.

For convenience, we introduce the transverse momentum probability density in symmetric nuclear matter,

$$f^*(k_\perp) = \frac{1}{4\pi^3 m_{K^+}^*} \int_0^{2\pi} d\phi \int_0^{m_{K^+}^*} \frac{dk^{*+} \mathcal{M}_0^{*2}}{k^{*+}(P^{*+} - k^{*+})} \times \Phi^{*2}(k^{*+}, \vec{k}_\perp; m_{K^+}^*, 0), \quad (28)$$

and from the integration of $f^*(k_\perp)$ we have the in-medium probability of the valence component in the kaon:

$$\eta^* = \int_0^\infty dk_\perp k_\perp f^*(k_\perp). \quad (29)$$

In the nuclear medium, the kaon decay constant, $f_{K^+}^*$, is defined as the matrix element of the partially conserved axial-vector current (PCAC) [4,5,9,33,51], and has the following expression [6]

$$P_\mu \langle 0(\rho)|A_i^\mu|K_j^*\rangle = im_{K^+}^{*2} f_{K^+}^* \delta_{ij}. \quad (30)$$

From the interaction Lagrangian density, Eq. (15), we obtain the in-medium decay constant, $f_{K^+}^*$, in terms of the valence component [5,6]

$$f_{K^+}^* = \sqrt{N_C} \int \frac{d^2 k_\perp dx^*}{(2\pi)^3} \frac{2[x^* m_{\bar{s}} + m_u^*(1-x^*)]}{x^*(1-x^*)} \times \Phi^*(x^*, k_\perp). \quad (31)$$

The normalization constant $N_C$ is obtained from the condition $F_{K^+}^*(0) = 1$. Then, the probability to find the kaon for the valence components state is $\eta^* = F_{K^+}^{*(WF)}(0)$, which is less than one at lower densities nuclear medium, similarly to the vacuum case [4,5]. In the present work, we also calculate the root-mean-square charge radius for the kaon, in the nuclear medium.

## IV. RESULTS

The pion model in the vacuum has two free parameters, the constituent up quark mass, $m_u = 220$ MeV [47,51–56], and the regulator mass, $m_R$, fixed in order to reproduce the correct experimental weak decay constant [18]. For the kaon in symmetric nuclear matter, we have the up quark mass in the medium, $m_u^*$, and the strange (antiquark) mass, $m_{\bar{s}}^* = m_{\bar{s}} = 508$ MeV, together with the regulator mass $m_R = 600$ MeV, as shown in Table II. The $m_R$ value is determined to fit the weak decay constant, Eq. (31), but the vacuum case [4]. The model gives, for the vacuum case, $f_{K^+} = 109.03$ MeV, which is within the experimental value of $f_{K^+}^{exp} = 110.096(3)$ MeV [18] (see Table III).

TABLE II. Parameters for the kaon in symmetric nuclear matter with the QMC model. In vacuum $R_K = 0.382$ fm and $Z_K = 3.920$ [see Eq. (9)].

| $\rho/\rho_0$ | 0.00 | 0.25 | 0.50 | 0.75 | 1.00 | 1.50 | 2.00 | 2.50 | 3.00 |
|---|---|---|---|---|---|---|---|---|---|
| $m_K$ [MeV] | 493.7 | 472.1 | 452.7 | 435.5 | 420.1 | 394.1 | 373.4 | 356.6 | 342.8 |
| $m_u$ [MeV] | 220 | 179.9 | 143.3 | 109.8 | 79.6 | 27.3 | −15.7 | −51.6 | −81.7 |
| $V_\sigma^u$ [MeV] | 0.00 | 29.2 | 58.4 | 87.6 | 116.9 | 175.3 | 233.7 | 292.2 | 350.6 |
| $m_{\bar{s}}$ [MeV] | | | | | 508.0 | | | | |
| $m_R$ [MeV] | | | | | 600.0 | | | | |

TABLE III. Summary of in-medium kaon properties calculated. The $\eta^*$ is calculated via Eq. (29), and gives the probability of the valence component in the kaon for the quark antiquark. The relation $\langle r_{K^+}^{*2}\rangle^{1/2} \cdot f_{K^+}^*$ is related to the Tarrach's theorem [57]. Parameters used in the calculations are $m_{\bar{s}} = 508$ MeV and $m_R = 600$ MeV.

| $\rho/\rho_0$ | $m_u$ [MeV] | $\langle r_{K^+}^{*2}\rangle^{1/2}$ [fm] | $f_{K^+}^*/f_{K^+}$ | $f_{K^+}^*$ [MeV] | $\eta^*$ | $\langle r_{K^+}^{*2}\rangle^{1/2} \cdot f_{K^+}^*$ |
|---|---|---|---|---|---|---|
| 0.00 | 220 | 0.712 | 1.000 | 109.01 | 0.711 | 0.388 |
| 0.25 | 179.925 | 0.792 | 0.952 | 103.82 | 0.752 | 0.411 |
| 0.50 | 143.261 | 0.896 | 0.906 | 98.79 | 0.802 | 0.438 |
| 0.75 | 109.890 | 1.024 | 0.863 | 94.03 | 0.864 | 0.470 |
| 1.00 | 79.619 | 1.182 | 0.822 | 89.66 | 0.943 | 0.507 |
| 1.50 | 27.346 | 1.661 | 0.758 | 82.60 | 1.161 | 0.583 |
| 2.00 | −15.725 | 1.746 | 0.717 | 78.20 | 1.346 | 0.545 |
| 2.50 | −51.558 | 1.278 | 0.682 | 74.32 | 1.366 | 0.439 |
| 3.00 | −81.727 | 1.101 | 0.647 | 70.49 | 1.375 | 0.364 |
| Exp.[PDG] [18] | $m_K = 493.7$ MeV | $0.560 \pm 0.031$ | | 110.096(3) | | 0.313 |

The electromagnetic root-mean-square charge radius, $\langle r_{K^+}^{*2}\rangle^{1/2}$, of the kaon is calculated from the elastic electromagnetic form factor, from the following expression [4,5],

$$\langle r_{K^+}^{*2}\rangle = -6\frac{\partial}{\partial q^2}F_{K^+}^*(q^2)|_{q^2\simeq 0}. \quad (32)$$

The result obtained for the kaon electromagnetic radius in vacuum with the present model, is $\langle r_{K^+}^{*2}\rangle^{1/2} = 0.712$ fm [4,33], to be compared with the experimental value $0.560 \pm 0.031$ fm [18], with an error of 21.3%. We evaluate numerically the derivative, and $|F_{K^+}^*(q^2)|$ varies fast around $q^2 = 0$ for all chosen nuclear densities as well as in vacuum. The values cited in this work are all evaluated at $Q^2 = -q^2 = 0.001$ (GeV/c)$^2$, where the stability of the form factor has been checked.

In Figs. 7 and 8, the $Q^2$ dependence of the kaon elastic electromagnetic form factor squared is shown in symmetric nuclear matter with the light-front model [4,5], for nine nuclear densities. Specifically in Fig. 8 we multiply the form factor by $Q^2$ to better analyze. The model result is compared with the experimental data in the vacuum [15,16].

The results for the model in the vacuum, reproduce well the experimental data for vacuum [15,16] and also agree with other models [7,8].

As the nuclear density increases, the absolute value of the form factor $|F_{K^+}^*(q^2)|$ decreases faster than in vacuum, and the electromagnetic charge radius for the kaon increase up to densities $\rho/\rho_0$ close to 1.8, where the mass of the quark u is positive, but for higher densities the behavior begins to change. This leads to a larger kaon charge radius in nuclear matter with increasing density, as shown in Fig. 9. In terms of the quark mass in the medium, $m_u^*$, the kaon electromagnetic radius decreases, like the pion case, consistent with the decreasing of the quark effective mass in the nuclear medium as shown in Fig. 10 (see the Table III).

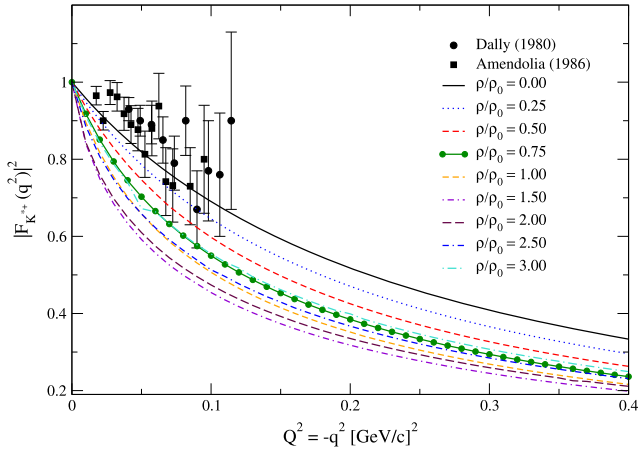

FIG. 7. Kaon electromagnetic form factor in symmetric nuclear matter for five nuclear densities as function of $Q^2 = -q^2$. Experimental data in vacuum are from Refs. [15,16].

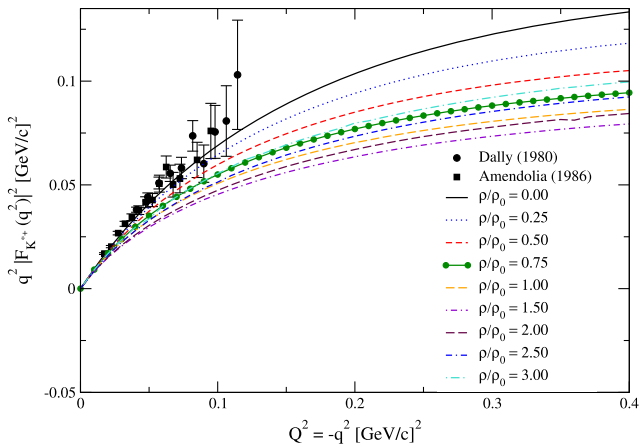

FIG. 8. The kaon electromagnetic form factor multiplied by $Q^2 = -q^2$.

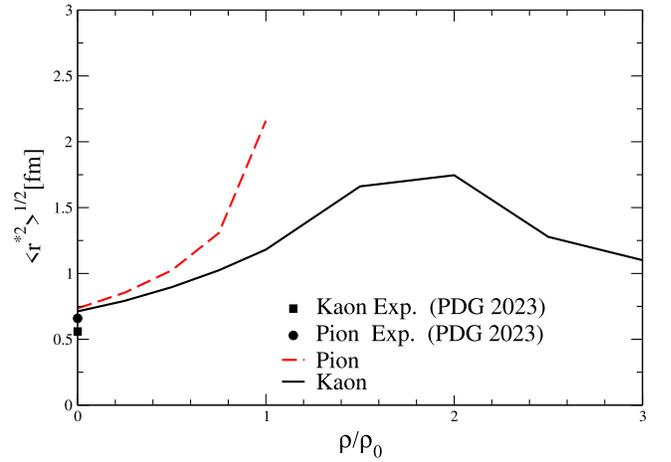

FIG. 9. The kaon and pion electromagnetic root-mean-square charge radius as function of the nuclear density.

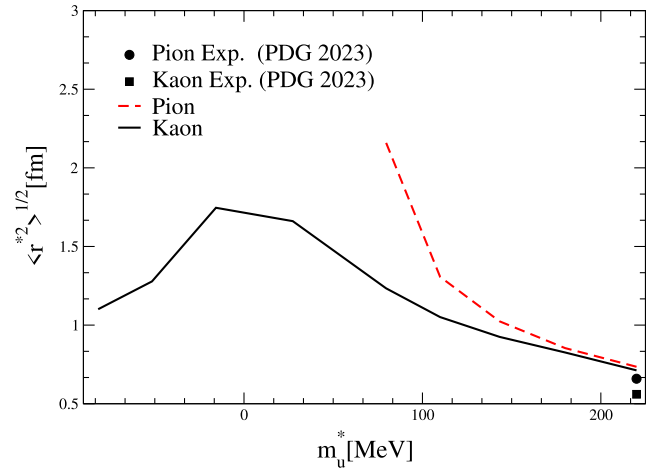

FIG. 10. The electromagnetic root-mean-square charge radius, $\langle r_{K^+}^*\rangle^{1/2}$, as a function of the quark mass, $m_u^*$.

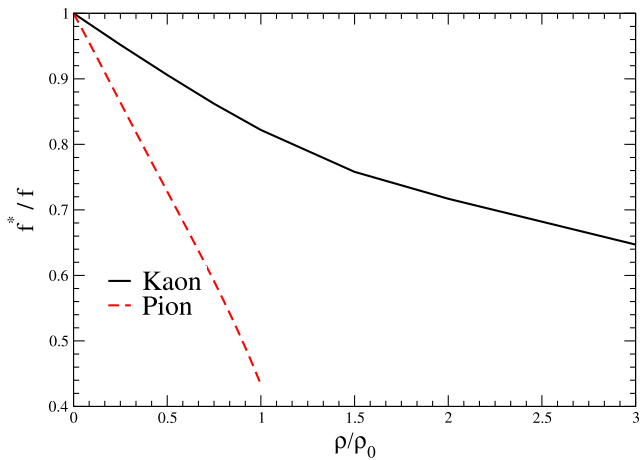

FIG. 11. Ratios the in-medium to vacuum electroweak decay constant, for the mesons pion and kaon, with the light-front time component, as function of the nuclear density.

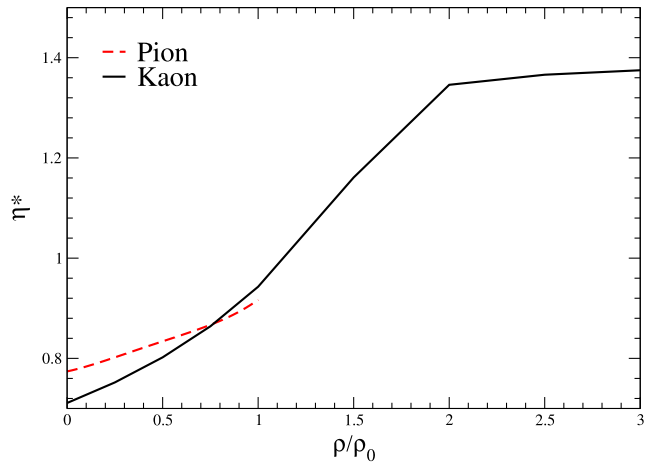

FIG. 12. In-medium valence probability $\eta^*$, as function of the nuclear density.

The decrease in the quark mass kinematically allows for the quarks to move in a larger space region and the quark-antiquark binding energy becomes smaller, i.e., the kaon is less bound which results in an increase of the charge radius. This can be demonstrated with the validity of Tarrach's theorem in symmetric nuclear matter described by approximately constant values of the relation $\langle r_{K^+}^{*2}\rangle^{1/2} \cdot f_{K^+}^*$ as shown in the last column of Table III. It is possible to see this theorem is satisfied also in other works in literature for pion and kaon in vacuum [4,5,33,58,59]. In Fig. 11, we show the ratios of the in-medium to vacuum kaon weak decay constant for the kaon in symmetric nuclear matter, $f_{K^+}^*/f_{K^+}$, versus nuclear density, associated with the light-front time component. The results show that $f_{K^+}^*$ and $f_{K^+}^*/f_{K^+}$ decreases as nuclear density increases (see Table III).

For the pion case, the bound state mass does not change so much in the nuclear medium, about ≈3.0 MeV. That is consistent with the empirical findings supported by the experiments [60], which yield $(f_{\pi^+}^*/f_{\pi^+})^2 = 0.64$ (associated with the time component) at density $\rho = 0.17$ fm$^{-3}$. In the case of the pion, the present model was exploited in the Ref. [6], producing a larger reduction. The similar behavior is found also, for the kaon in the present work.

The kaon properties in symmetric nuclear matter are summarized in Table II. In the table, we show that the kaon bound state mass decreases with the nuclear density, which is a feature different from the pion case, where the bound state effective mass is almost constant in symmetric nuclear matter [6]. Also, the quark mass $m_u^*$ decreases with the nuclear density, and this is the same for the pion and kaon [6,37]. The potential, related with the elastic electromagnetic form factor, Eq. (24), is in the fourth row of Table II. The third, fourth, fifth, and sixth columns in Table III are the in-medium quantities calculated with the present model respectively: kaon electromagnetic root-mean-square charge radius, the ratio in-medium to vacuum electroweak decay constant, the kaon electroweak decay constant and the valence quark probability $\eta^*$. In Fig. 12 one can see that $\eta^*$ grows for larger values of nuclear densities, note that for $\rho > 1.2\rho_0, \eta^* > 1$, it is possible to see in Eqs. (28) and (29), since $m_K^* > m_K$ and $\eta^* \propto 1/m_K^*$. This may indicate that $K^+$ becomes a quasibound state and not anymore the pure $|u\bar{s}\rangle$ composite state, but it contains more valence quarks.

As nuclear density increases, the probability of the valence component in the kaon is enhanced, which is again the effect of the decreasing of the effective quark mass. This makes the light $u$ quark to move more freely inside the kaon. This effect has the same origin as the increase of the kaon root-mean-square charge radius in symmetric nuclear matter discussed above.

The results obtained in the Figs. 13–16 through the Eqs. (26) and (27) show the behavior of the electromagnetic form factor in vacuum and in medium, the contributions of the quark $u$, $F^*_{K^+(u)}$, and the strange antiquark, $F^*_{K^+(\bar{s})}$ with the values of $e_u = 2/3$ and $e_{\bar{s}} = 1/3$.

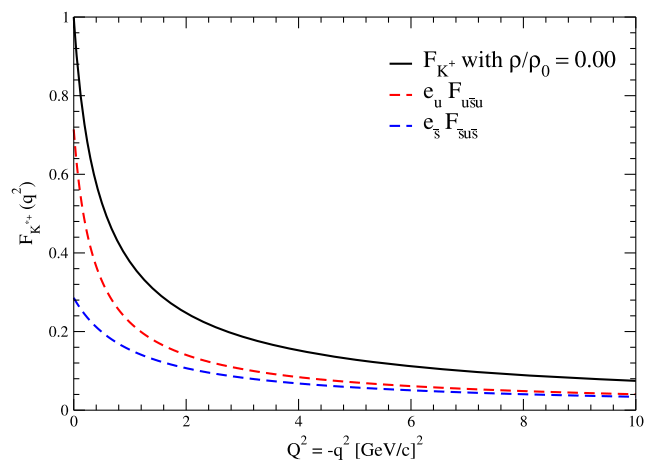

FIG. 13. $F_{K^+}(Q^2)$ decomposition of quark contributions for the vacuum case.

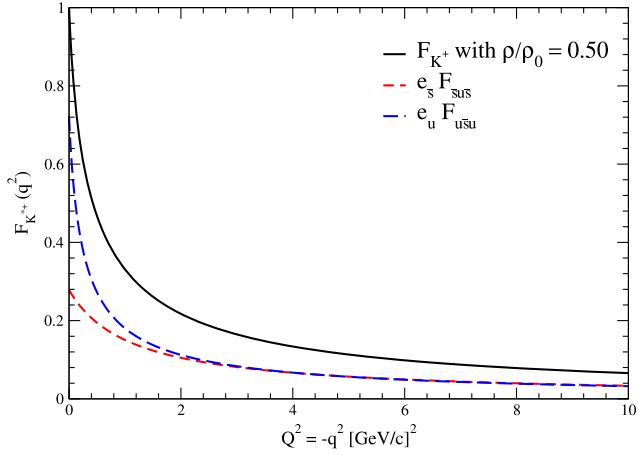

FIG. 14. $F_{K^+}(Q^2)$ decomposition of quark contributions for $\rho/\rho_0 = 0.50$.

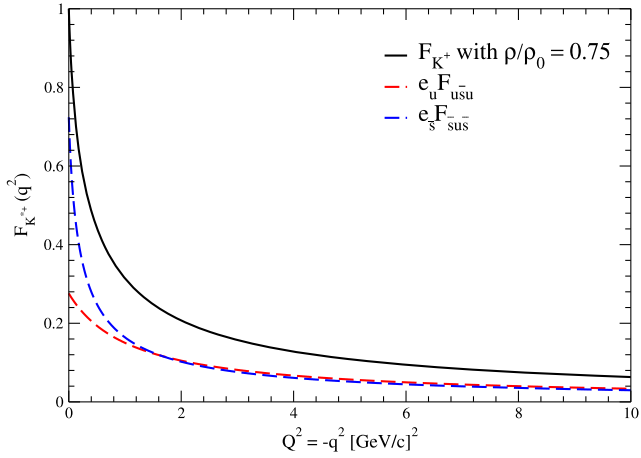

FIG. 15. $F_{K^+}(Q^2)$ decomposition of quark contributions for $\rho/\rho_0 = 0.75$.

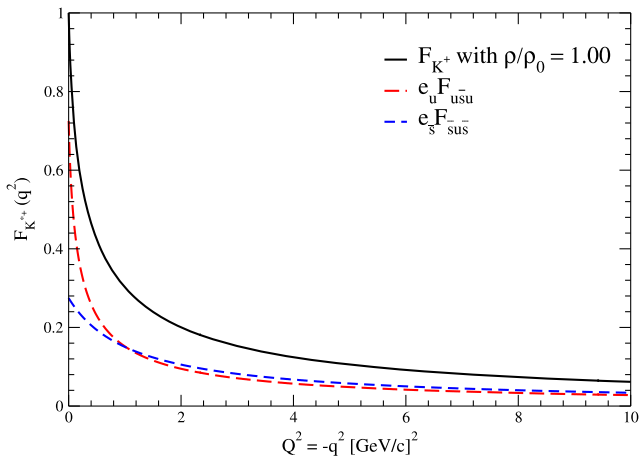

FIG. 16. $F_{K^+}(Q^2)$ decomposition of quark contributions for $\rho/\rho_0 = 1.00$.

TABLE IV. Decomposition of quark contributions for the kaon electromagnetic elastic form factor at $Q^2 = 0$, with the parameters, $m_{\bar{s}} = 508$ MeV and $m_R = 600$ MeV. in nuclear medium.

| $\rho/\rho_0$ | $e_{\bar{s}} F_{\bar{s}u\bar{s}}(0)$ | $e_u F_{u\bar{s}u}(0)$ | $F_{K^+}(0)$ |
|---|---|---|---|
| 0.0 | 0.2858 | 0.7142 | 1.0 |
| 0.5 | 0.2783 | 0.7213 | 1.0 |
| 0.75 | 0.2760 | 0.7240 | 1.0 |
| 1.00 | 0.2749 | 0.7251 | 1.0 |

Analyzing these figures, we note that at $Q^2 = 0$, we have that the contribution of the quark $u$ is around 66.67%, while that of the antiquark $\bar{s}$ is 33.33%, see Table IV, and for larger transferred momentum, equality in slopes does not lead to equality in the results obtained for the total radius and its partial contributions.

In Fig. 13, we show the behavior of the electromagnetic form factor in vacuum $\rho/\rho_0 = 0.00$, and contributions of the u quark and the strange antiquark. It is observed that for values of transferred momentum-squared close to $Q^2 = 8$ [GeV/c]$^2$, the contributions are close. Whereas in Fig. 14 we show the behavior of the electromagnetic form factor at $\rho/\rho_0 = 0.50$ for values of transferred momentum-squared close to $Q^2 = 2.2$ [GeV/c]$^2$. Approximately, the curves of the partial contributions are close again. But in Fig. 15 the behavior of the electromagnetic form factor at $\rho/\rho_0 = 0.75$ are inverted for $Q^2 = 1.8$ [GeV/c]$^2$, evidencing the greater contribution of the u quark. This also happens at $\rho/\rho_0 = 1.00$ for values of transferred momentum close to $Q^2 = 1$ [GeV/c]$^2$.

## V. SUMMARY AND CONCLUSIONS

We have explored the modifications of the kaon properties in symmetric nuclear matter based on the light-front constituent quark model [4,5,30,31], plus quark-meson coupling (QMC) model [6,41–43]. The strategy adopted here reproduces quite well the experimental data in vacuum, such as the electromagnetic elastic form factor, and other observables, for example, electromagnetic root-mean-square charge radius and the electroweak decay constant, where we use the light-front plus component of the electromagnetic current [6,37]. In order to incorporate the nuclear many-body effects on an equal footing, i.e., with the quark degrees of freedom, we have employed the QMC model. We have used the in-medium quark properties obtained in the QMC model as input for the constituent up and strange quarks. For the kaon the in-medium kaon mass is also calculated by the QMC model and the vector potential at normal nuclear matter density is slightly repulsive. Those properties are summarized in Table II for the effective kaon mass and in Table III for its electroweak properties.

The kaon properties calculated in symmetric nuclear matter are the elastic electromagnetic form factor, root-mean-square charge radius, the weak decay constant, and the valence probability of $q\bar{q}$, for the kaon in-medium. Like the pion case [4,6], the results indicate a rapid decrease for electromagnetic elastic form factor in-medium with increasing the nuclear density, and as a result an increase of the kaon root-mean-square charge radius.

We have also computed the in-medium kaon decay constant, which is associated with the light-front time component. The decay constant decreases as nuclear density increases, which is consistent with the analysis of the pion case already obtained in the literature.

The corresponding ratio, $f^*_{K^+}/f_{K^+}$, obtained in the present approach is decreased, or equivalently, the reduction of $f^*_{K^+}$ is larger. However, we must state that there is an uncertainty in the change of kaon effective mass in the medium, from which the value of the kaon decay constant is calculated. Regarding the valence quark probability in $K^+$, our result shows that this probability increases with the increase of density. We understand this effect in terms of the decreasing of the effective constituent light up quark mass in kaon, which allows for a larger kinematic distribution within the kaon and, in turn, results in an increase in the probability of valence quark probability. The same reasoning applies to the increase of the kaon charge radius. Our next step may be extended the present approach to the $B$, $D$, and $\omega$ mesons in vacuum and in nuclear medium.

## ACKNOWLEDGMENTS

This work was financed in part by the Fundação de Amparo à Pesquisa do Estado de São Paulo (FAPESP), Brazil, No. 2019/02923-5 and the Conselho Nacional de Desenvolvimento Científico e Tecnológico (CNPq) Process, No. 307131/2020-3, with J. P. B. C.M. The work of K. T. was supported by the CNPq Process, No. 426150/2018-0, and No. 304199/2022-2, and FAPESP Process, No. 2019/00763-0 and No. 2023/07313-6. The work of T. F. acknowledge de partial support by CNPq, under the grant No. 306834/2022-7, and by FAPESP under the Thematic grant No. 2019/07767-1 and W.d.P. acknowledges the partial support of CNPq under Grants No. 313030/2021-9 and the partial support of Coordenação de Aperfeiçoamento de Pessoal de Nível Superior (CAPES) under Grant No. 88881.309870/2018-01. The work was also part of the projects, Instituto Nacional de Ciência e Tecnologia—Nuclear Physics and Applications (INCT-FNA), Brazil, Process. No. 464898/2014-5.